\documentclass[aps,prl,twocolumn,showpacs,superscriptaddress,groupedaddress]{revtex4}  
\usepackage{graphicx}
\usepackage{hyperref}
\usepackage{dsfont,amsmath,amssymb,color,units,pstricks}
\usepackage{amsfonts}
\newcommand\be{\begin{equation}}
\newcommand\bea{\begin{eqnarray} \nonumber }
\newcommand\ee{\end{equation}}
\newcommand\eea{\end{eqnarray}}

\begin{document}

\title{Slow decay of impact in equity markets}

\author{X. Brokmann, E. S\'eri\'e, J. Kockelkoren, J.-P. Bouchaud}
\affiliation{Capital Fund Management - 23, rue de l'Universit\'e 75007 Paris - France}


\date{\today}

\begin{abstract}
Using a proprietary dataset of meta-orders and prediction signals, and assuming a quasi-linear impact model, we deconvolve market impact from past correlated trades and a predictable return component 
to elicit the temporal dependence of the market impact of a single daily meta-order, over a ten day horizon in various equity markets. We find that the impact of single meta-orders 
is to a first approximation universal and slowly decays to zero (or to a small value), possibly as a power-law. We show that auto-correlated order-flows and trade information contents fully accounts 
for the apparent plateau observed in the raw data. We discuss the possible bias introduced by the quasi-linear assumption.
\end{abstract}


\maketitle



\section{Introduction}

One of the most fundamental question in financial economics is the shape and nature of price impact, to wit: how 
buying (/selling) a quantity $Q$ of shares (or contracts on future markets, or any other traded asset) will 
move the price, in particular when this quantity $Q$ is executed incrementally over time \cite{Kyle:1985,Bouchaud:2008}. There is now a 
remarkable consensus that the impact of such a {\it meta-order} -- measured as the average difference between the final price and the initial
price times the sign of the order -- has an approximate square-root dependence on $Q$ (see e.g. \cite{Torre:1997,Almgren:2005,Moro2009,Toth:2011,Bladon2013,Kyle:2012,Iacopo:2013,Zarinelli:2014}). 
More precisely, a very large number of independent quantitative measures of price impact have concluded that to a first approximation, and independently 
of the asset class, time period, style of trading and micro-structure peculiarities, one has (see Fig. \ref{Fig1}):  
\begin{equation}\label{Yformula}
\left\langle \epsilon \cdot \frac{p(T) - p(0)}{p(0)} \right\rangle \approx Y_0 \sigma \left(\frac{Q}{V}\right)^\delta \equiv \theta,
\end{equation} 
where $\epsilon = \pm 1$ for buy (sell) orders, $p(0)$ is the price at the start of the execution period and $p(T)$ the price at the end of the execution period.
Here and below, $\langle ... \rangle$ denote averaging. In the right hand side of the equation, $\sigma$ and $V$ are, respectively, the volatility of the traded instrument and the total traded volume over the order execution time scale $T$, $Y_0$ a dimensionless constant of order 
unity, and $\delta$ is the impact exponent in the range $[0.4, 0.7]$. Theoretical models allowing one to understand the mechanisms underlying this strongly concave, non-intuitive impact function, have been put 
forward in \cite{Toth:2011, Farmer:2011, Iacopo:2013, Iacopo:2014} (on a related topic, see also \cite{Damian:2014}).

Still, whereas the description of the contemporaneous impact of orders in terms of Eq. (\ref{Yformula}) is by now well accepted, the fate of this impact {\it after the order has been completed} is still not settled and 
is a matter of debate. On general grounds, one expects this impact to relax from its peak value, since upon conditioning on the fact that the extra buyer (seller) is no longer active, the subsequent flow is biased in the
opposite direction. This is clearly the case in the ``latent order book'' model of T\'oth et al. \cite{Toth:2011, Iacopo:2013}, but has also been argued on theoretical grounds in \cite{Farmer:2011} (see also \cite{Donier:2012}). Based on fair pricing (no arbitrage) arguments, the authors of \cite{Farmer:2011} propose that 
the asymptotic value of the impact (long after the order has been completed) should be a fraction of the peak value, in such a way that the average price paid by the buyer (seller) is equal to the long term value. This 
implies zero average profit both for the liquidity provider and for the trader. For a concave impact and a constant rate of execution, it is easy to derive from Eq. (\ref{Yformula}) that the average paid price is $\theta/(1+\delta)$ above the initial price $p(0)$. Therefore, the long-time plateau value of the impact is predicted to be $2/3$ of the peak value when $\delta=1/2$. This prediction seems to be borne out by two empirical papers \cite{Moro2009,Bershova2013}, while a third study \cite{Gomes2013} reports a more complex picture, where ``informed'' trades and ``uninformed'' trades lead to a very different impact relaxation pattern. In the former case, the
impact seems to relax towards the predicted $2/3$ value, while in the latter case, impact appears to relax all the way to zero. The authors of \cite{Gomes2013} argue that uninformed trades should have no long term impact, and 
that the long-term impact of informed trades should reflect, on average, the information content of these trades. Incidentally, both papers \cite{Bershova2013, Gomes2013} confirm once again the validity of square-root impact law, Eq. (\ref{Yformula}).

While we agree with the general ideas expressed in \cite{Gomes2013}, there are several issues that need to be clarified. First, from a theoretical point of view -- if some trades are uninformed and others are informed, the 
simple fair pricing argument of \cite{Farmer:2011} becomes more subtle since liquidity providers can make money out of ``noise traders'', as in Kyle's model \cite{Kyle:1985}. Second, as recognized in \cite{Gomes2013}, the notion of informed trades is at best ambiguous and needs to be supplemented by a time horizon, over which the prediction is supposed to be realized. If buying a stock is with the objective of very short-term gains, on the same scale as the measurement of impact relaxation, then the distinction between informed and uninformed trades may be warranted. But if the future gain is over 
several months while impact is measured over a few hours to a few days (beyond which the noise level becomes insuperable), then clearly the distinction is moot. Third, and perhaps most importantly, there is a methodological 
problem with the way the data has been analyzed in all previous work on the subject. The point is that proprietary order flows tend to be highly auto-correlated in time (due to the slicing of large buy/sell orders over several days or
sometimes weeks). Price dynamics seen after completion of an intra-day meta-order therefore reflects {\it both} its own decaying impact and the impact dynamics from all previous orders as well. This requires some deconvolution 
treatment that turns out to be crucial to properly address the question of impact decay -- as the empirical and theoretical analysis below clearly demonstrates.   

In this Letter, we analyze the impact of CFM's proprietary trades on equity markets. We show that our data is, after deconvolution, compatible with a price model 
where the ``mechanical'' impact of meta-orders decays all the way to zero, superimposed with some (investor dependent) predictability pattern. 
In other words, we posit that the {\it average} price profile after the completion, at time $t$, of a single meta-order, of volume $Q$ and predictability $\alpha$, is the sum of two fundamentally different contributions:
\be\label{linear-model}
\mathbb{E}_t[p(t > T)] \approx p(0) \big[1 +  \theta(Q) I(t-T) + \alpha H(t)\big].
\ee
In the above equation, $\theta(Q) \equiv \epsilon(Q) Y_0  \sigma \left[{Q}/{V}\right]^\delta$ is the instantaneous impact of the trade, $I(\tau)$ is a ``propagator'' describing the mechanical part of the impact \cite{Bouchaud:2004}, which decreases from $I(0)=1$ to $I(\infty) \approx 0$. The predictor amplitude $\alpha$ (possibly equal to zero for uninformed trades) 
is the expected long term price profile that motivated the trade at time $t=0$ 
and $H(t)$ describes the way this prediction realizes itself over time. For example, for an Ornstein-Uhlenbeck predictor, $H(t) = 1 - \exp(-\Gamma t)$ for a prediction on a horizon $\sim \Gamma^{-1}$ (see below). 
For uninformed trades or for bets on very long horizons such that $\Gamma t \ll 1$, one can set 
$\alpha = 0$. This, we believe, allows one to stitch together various aspects of the literature in a coherent way. 
However, as we shall argue in the conclusion, the model defined by Eq. (\ref{linear-model}) is not entirely 
satisfactory, since it is quasi-linear (different meta-orders add up linearly) whereas the concavity of each single meta-order [Eq. (\ref{Yformula})] indicates strong non additive behaviour, at least on short time scales.

\begin{figure}[b]
\includegraphics[scale=0.3]{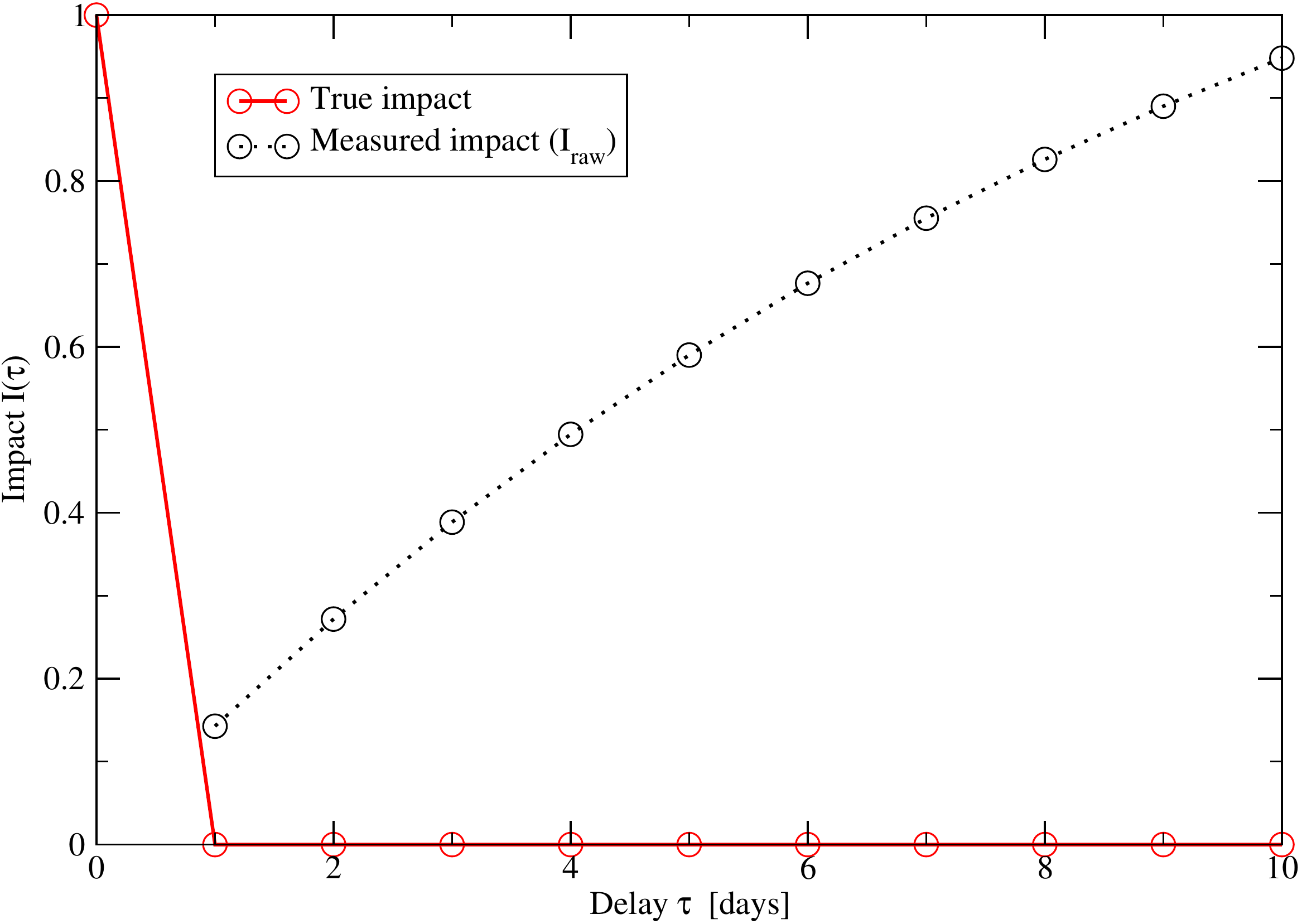}
\caption{Illustration of the difference between the ``true'' impact $I(\tau)$ of an isolated order (here chosen to be a $\delta$-function in time) and the 
apparent impact $I_\mathrm{raw}(\tau)$ of a series of correlated trades, resulting from optimizing the expected gain of a strategy based on an 
Ornstein-Uhlenbeck predictor, with $\Gamma=0.1$ and $z=0.5$.\label{Fig0}}
\end{figure}

\section{An illustrative toy-model}

Let us first present a toy-model that highlights the difference between the ``true'' impact $I(\tau)$ and the impact of a sequence of correlated trades, resulting from an investor who attempts to optimize his 
gains. We consider an idealized market where impact is {\it linear} and decays instantaneously, i.e. the true impact is  $I(\tau)=\gamma \delta(\tau)$. 
We imagine that the investor has an Ornstein-Uhlenbeck 
signal $s(t)$ with correlation time $\Gamma^{-1}$, such that:
\be
\langle s(\tau) s(0) \rangle = e^{-\Gamma |\tau|}.
\ee
The prediction based on this signal is that the price change between $t$ and $t+{\rm d} t$ is:
\be
\langle p_\mathrm{d}(t+{\rm d} t) - p_\mathrm{d}(t) | s(t) \rangle = {\cal G} s(t) {\rm d} t,
\ee
while the integrated expected gain up to time $t+\tau$ is:
\be
\langle p_\mathrm{d}(t+ \tau) - p_\mathrm{d}(t) | s(t) \rangle = s(t) \frac{{\cal G} }{\Gamma} \left[1 - \exp(-\Gamma \tau)\right].
\ee
The investor wants to optimise the total gain, impact costs included, for a fixed level of risk $R^2$. This is implemented (in continuous time) as 
the maximisation of:
\be
{\cal L} = \frac1T \left[ \int_0^T {\rm d}t \pi(t) {\cal G} s(t) - \frac{\gamma}{2} \int_0^T {\rm d}t \dot \pi(t)^2 - \lambda \int_0^T {\rm d}t \pi(t)^2 \right]
\ee
where $\pi(t)$ is the position (in \$) at time $t$ and $\gamma$ is the impact coefficient: trading at speed $\dot \pi$ pushes the price 
instantaneously by an amount $\gamma \dot \pi$. The average price paid in the process is $\frac12 \gamma \dot \pi$.
$\lambda$ is the Lagrange multiplier allowing one to fix the average risk at a chosen value $R^2$. We assume that
$T \to \infty$ in the following and forget about the bounds on $t$.

The above quadratic optimisation problem can easily be solved, see e.g. \cite{Emeric,Galeanu}. The solution of this problem is to fix the position $\pi(t)$ 
proportional to an exponential moving average of the predictor:
\be
\pi(t) = \phi_0 \int_{-\infty}^t {\rm d}t' e^{-\omega(t-t')} s(t'),\quad \phi_0 = \frac{{\cal G}}{\gamma (\omega + \Gamma)}
\ee
where the averaging frequency $\omega$ is set by the following equation:
\be
z (1+z)^3 = \frac{\sigma^2 {\cal G}^2}{\gamma^2 R^2 \Gamma^4}; \qquad \omega = z \Gamma.
\ee
As expected, $z \to \infty$ when $\gamma \to 0$, i.e. the position tracks instantaneously the signal. As the friction becomes higher however, the 
signal has to be slowed down in order to extract a positive gain from the signal. 

Now, we want to compute the {\it measured impact} induced by this optimal trading policy. More precisely, we will define the ``raw'' impact 
$I_\mathrm{raw}(\tau)$, normalized by
the average price impact due to trading at the end of the first trading period (i.e. at $t + {\rm d}t$, just before the ``true'' impact decays back to zero), as:
\be
\langle (p_\mathrm{d}(t+ \tau) - p_\mathrm{d}(t)) \dot \pi(t) \rangle = \gamma \langle \dot \pi(t)^2 \rangle \times I_\mathrm{raw}(\tau), \qquad (\tau > 0).
\ee
This is the impact one would measure by averaging the average price profile over all the trades of the investor, without any deconvolution (hence the term ``raw''). 
It is easy to show that:
\be
\langle (p_\mathrm{d}(t+ \tau) - p_\mathrm{d}(t)) \dot \pi(t) \rangle=  \gamma \phi_0^2  \left[1 - \exp(-\Gamma \tau)\right]
\ee
which has to be normalized by:
\be
\gamma \langle \dot \pi(t)^2 \rangle = \frac{1}{(z+1)} \gamma \phi_0^2,  
\ee
finally leading to (see  Fig. \ref{Fig0}):
\be
I_\mathrm{raw}(\tau) = (1+z) \left[1 - \exp(-\Gamma \tau)\right].
\ee
This calculation reveals that although $I(\tau)$ is a $\delta$-function in time, $I_\mathrm{raw}(\tau)$ increases from zero to a non universal plateau value $1+z$ 
which depends continuously 
on the a-dimensional variable $\gamma R \Gamma^2/\sigma {\cal G}$ that measures the relative strength of the impact costs and the expected gains \footnote{Note 
in particular that the ``fair value'' advocated in \cite{Farmer:2011} 
(corresponding, with the present normalisation, to $I=1/2$) does not seem to play any special role here.}.

\begin{figure}[b]
\includegraphics[scale=1]{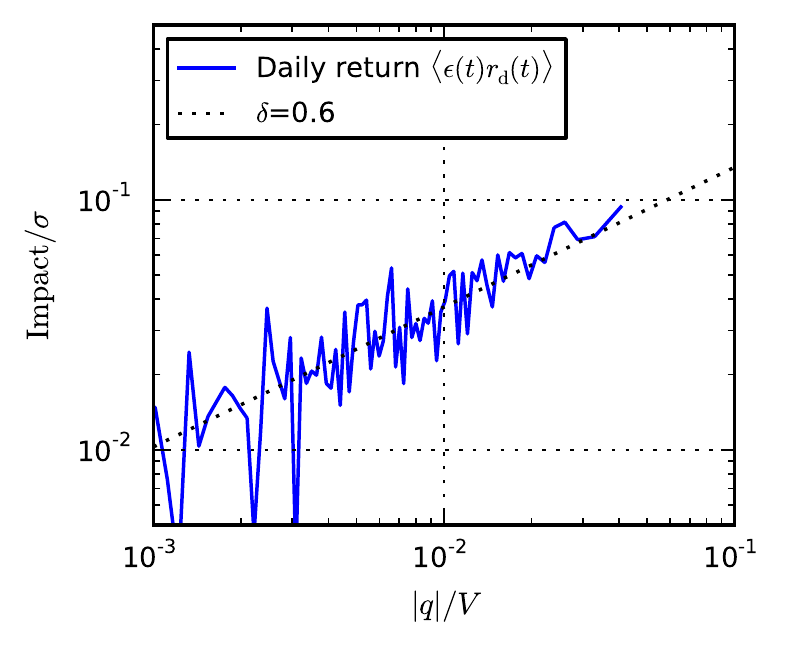}
\caption{Impact as a function of the volume fraction $|q|/V$ of traded meta-orders on equity markets. The line is a moving average through a cloud of points corresponding to
individual meta-orders (not shown). 
The dependence is concave and follows a power-law scaling of exponent $\delta=0.6$ over the whole range of $|q|/V$.
This value of $\delta$ turns out to be exactly the same as the one obtained by Almgren et al. \cite{Almgren:2005}.\label{Fig1}}
\end{figure}


\section{Empirical data and impact decay}

The above toy-model clearly shows that in order to measure the ``true impact'' $I(\tau)$, a careful deconvolution of the raw impact must be performed.
Our empirical analysis is based on a proprietary CFM dataset, consisting in the exhaustive list of 1.6 million meta-orders, together with the daily signals that
lead to these trades, over three full years (2011-2013) in a variety of equity markets (Europe, US, Japan and Australia). In this dataset, each transaction 
derives from a daily trading decision, resulting in a sequence of daily meta-orders of signed quantities $\{q(0),q(1),\ldots,q(t),q(t+1),\ldots\}$ (where $q(t) := \epsilon(t) Q(t)$), 
starting at decision (midpoint) prices $\{p_\mathrm{d}(0),p_\mathrm{d}(1),\ldots,p_\mathrm{d}(t),p_\mathrm{d}(t+1),\ldots\}$ and executed over the course of the day at average prices 
$\{p_\mathrm{x}(0),p_\mathrm{x}(1),\ldots,p_\mathrm{x}(t),p_\mathrm{x}(t+1),\ldots\}$. Typical volume fractions are in the range $Q/V \sim 0.1 - 5 \%$ 
(see Fig. \ref{Fig1}). 
Note that orders, when started, are always (nearly) 
fully executed, with average execution rate exceeding 99\%. There is therefore no selection bias in the execution price of our trades. 
The daily signals $\{\pi(0),\pi(1),\ldots,\pi(t),\pi(t+1),\ldots\}$ correspond to CFM's best prediction of future prices, and are a combination of a variety of 
medium to long term technical and fundamental indicators, and with a prediction horizon $\Gamma^{-1}$ ranging from 10 to 100 days.  
The signals $\pi(t)$ are computed just before the execution of the meta-order and are normalized to 
have unit variance. The signals are furthermore made market neutral so all results reported below are, to first order, immune to market directional effects.  

Due to price impact, the execution prices $p_\mathrm{x}(t)$ are on average higher (lower) than the initial market midpoint $p_\mathrm{d}(t)$ when the buy (sell) meta-order started, 
causing both the ``strike slippage'' $r_\mathrm{x}(t)=(p_\mathrm{x}(t)-p_\mathrm{d}(t))/p_\mathrm{d}(t)$ and the price change $r_\mathrm{d}(t)=(p_\mathrm{d}(t+1)-p_\mathrm{d}(t))/p_\mathrm{d}(t)$ 
between two daily trading decisions to exhibit the expected square-root dependence on $Q/V$ (see Fig. \ref{Fig1}, where the fitting line corresponds to $\delta = 0.6$), 
in agreement with Eq. (\ref{Yformula}) and well in line with all previous research, see in particular \cite{Almgren:2005}. 

\begin{figure}[t]
\includegraphics[scale=1]{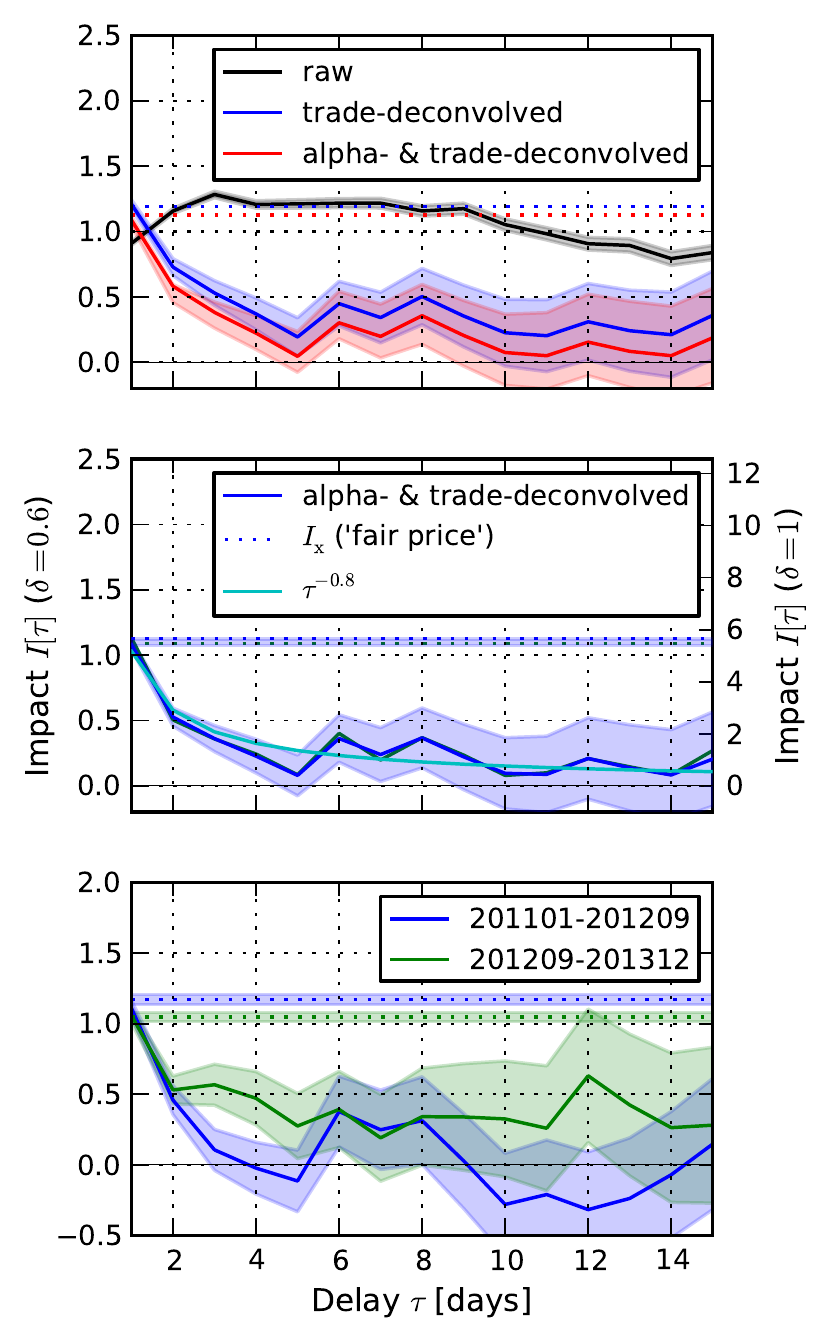}
\caption{Impact dynamics, obtained through the regression Eq. (\protect{\ref{rawFit}}), averaged over all trades, 
markets and periods with an equal weight, normalized by the execution price $I_\mathrm{x}$. The
shaded region corresponds to 2-$\sigma$ error bars. The lag is counted in business days.
a) Raw impact appears to reach a plateau for the raw data (black line) but true impact is found to be decaying (blue/red lines) once returns are corrected from the influence of past trades and predictability. 
The price reverts to its pre-trade value, or at least well below the average the execution price (horizontal dotted line at $y=1$). 
b) Impact deconvolved profiles for $\delta=0.6$ (green) and $\delta=1$ (blue), and simple power-law fit, yielding $I(\tau) \propto \tau^{-0.8}$.
c) Impact decay profiles as measured over two disjoint time periods. \label{Fig2}}
\end{figure}

We now turn to measuring the decay of impact over subsequent days. We first determine the raw average impact $I_\mathrm{raw}(\tau)$ through the following regression:
\begin{equation}
\label{rawFit}
\frac{p_\mathrm{d}(t+\tau)- p_\mathrm{d}(t)}{p_\mathrm{d}(t)}= I_\mathrm{raw}(\tau)\theta(t) + \xi(t),
\end{equation} 
where $\xi$ is a noise term, and again $\theta(t) \equiv Y_0\epsilon(t)\sigma [Q(t)/V]^\delta$. Note that by definition, $I_\mathrm{raw}(\tau)$ is normalized to 
unity for $\tau=1^-$, i.e. at the end of the meta-order of the first day (as for our toy model above). We also define, in the same units, the average cost $I_\mathrm{x}$ 
incurred by our execution, as caused by the impact of this order on $p_\mathrm{x}(t)$ and any other micro-structural effects: 
$r_\mathrm{x}(t)= I_\mathrm{x} \theta(t)$. 
The result of the regression Eq. (\ref{rawFit}) averaged over all markets and all periods is shown as the roughly constant black line in  Fig. \ref{Fig2}-a. Note that in this graph 
we have divided $I_\mathrm{raw}(\tau)$ by $I_\mathrm{x}$, in such a way that ``fair pricing'' should yield a horizontal line $y=1$. One sees that, 
in apparent agreement with the fair-price argument, the (rescaled) price indeed remains in the vicinity of the executed price $I_\mathrm{x}$ 
several days after the initial meta-order was completed .

However, the autocorrelation $\langle \epsilon(t)\epsilon(t+\tau)\rangle$ of the sign of the trades is found significantly positive over several days, 
obfuscating, as illustrated by our toy-model above, how much of the raw dynamics $I_\mathrm{raw}(\tau)$ actually reflects the impact of a single meta-order. In other words, the impact 
of the trades following the initial one keeps the impact $I_\mathrm{raw}(\tau)$ artificially high. Furthermore, since the trades are initiated in part by 
predictive signals, one should expect this signal to reveal itself through time and contribute to $I_\mathrm{raw}(\tau)$.

Instead of temporally aggregating successive meta-orders of the same sign as in \cite{Bershova2013,Gomes2013} 
(a non-causal procedure which, as such, induces conditioning artefacts), 
or using a somewhat ad-hoc correction to compensate for the trades occurring during the decay of impact \cite{Gomes2013}, we propose to 
deconvolve jointly past trades and predictions from the raw impact dynamics assuming a quasi-linear model, i.e. linearly regressing the daily returns $r_\mathrm{d}(t)$ on both past impacts 
$\{\theta(t),\theta(t-1),\ldots\}$ and return predictions $\{\pi(t), \pi(t-1),\ldots\}$, exploiting the ability of multiple least-square regression to properly account for all correlations among regressors:
\begin{equation}\label{MultiReg}
r_\mathrm{d}(t) = \sum_{t'=t-M}^t G_\theta(t-t') \theta(t') + \sum_{t'=t-M}^{t}G_\pi(t-t')\pi(t') + \xi(t),
\end{equation} 
where $M$ is a certain maximum lag that does not affect much our conclusions provided it is taken to be larger than $5$ days. In fact, $G_\pi(t-t')$ is found to be close to $G_\pi(0)\delta_{t,t'}$, as expected if the predictor is itself optimal. 
Instrumental aspects to successfully derive the kernels $G(\tau)$ include i) working on a \emph{complete} set of proprietary trades 
-- given the potential of incomplete datasets to exhibit detrimental bias in the trade autocorrelation estimation - and ii) performing the linear regression on market zones 
separately to ensure maximal homogeneity within the dataset - as required in any basic least-square deconvolution.

From the determination of $G_\theta(\tau)$, the ``true'' impact kernel $I(\tau)$ (stripped out of the contribution of both predictions and correlated trades) 
can be reconstructed as  $I(\tau)=\sum_{t'=0}^{\tau} G_\theta(t')$. The average returns during meta-order execution $r_\mathrm{x}(t)$ are treated similarly to compute 
the corresponding $\{G_{\theta,\mathrm{x}}(t')\}_{t'\in [0,M]}$, with $G_{\theta,\mathrm{x}}(0)$ providing an execution price proxy itself deconvolved from past trades and predictions. 

\begin{figure}[t]
\includegraphics[scale=1]{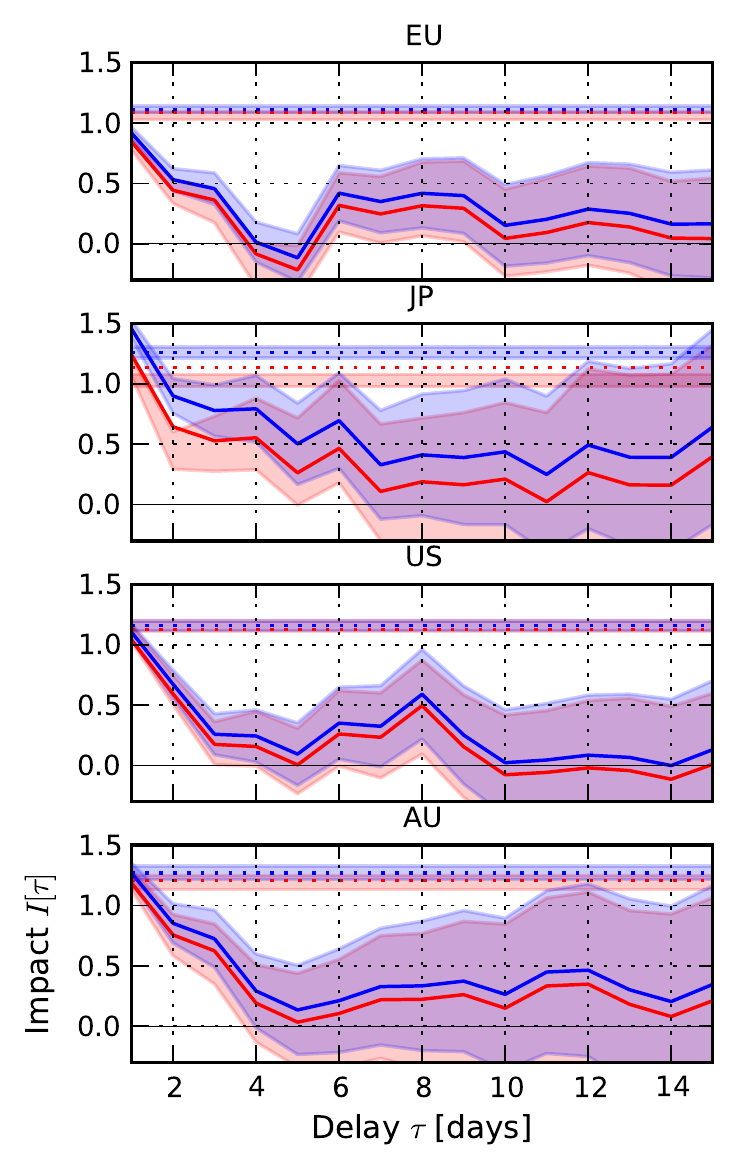}
\caption{Impact dynamics for different equity markets (here EU, US, Australia \& Japan), once deconvolved of the influence of past trades (red) and predictability (blue), 
and normalized by the average execution price $I_\mathrm{x}$.\label{Fig3}}
\end{figure}

Figure \ref{Fig2}-a illustrates the result of this approach to deconvolve $I_\mathrm{raw}$ from either past trades only (i.e. enforcing $G_\pi(t) \equiv 0$ in Eq. \ref{MultiReg}), 
or both past trades and predictors. One can see from Fig. \ref{Fig2}-a that the main effect in fact comes the correlation of meta-orders, which, when taken into account, 
leads to a significant decay of $I(\tau)$, much below the execution price $I_\mathrm{x}$ (blue line). Removing the 
prediction signal decreases further the impact, which decays to a very small value beyond 10 days or so (red line). The difference between the red and the blue line is however small, 
meaning that the average prediction strength of the signal $\pi(t)$ is small on short term horizons (in line with the fact that $\Gamma^{-1} \sim 10-100$ days).
The decay profiles are also independent on the exact value of the impact exponent $\delta$ that is used to 
define $\theta(t)$. Fig. \ref{Fig2}-b shows that the decay profile remains essentially unchanged for $\delta$ ranging from $0.5$ (square root impact) to $1$ (linear impact) -- providing strong evidence that the observed decay of
$I(\tau)$ is a genuine property of impact dynamics rather than a deconvolution artefact. A simple power-law fit suggests a slow decay $I(\tau) \sim \tau^{-0.8}$, similar to the decay of impact of individual trades, see 
\cite{Bouchaud:2004, Bouchaud:2005}. Deconvolved profiles appear reasonably stable both in time (Fig. \ref{Fig2}-c) and across equity markets (Fig. \ref{Fig3}). 
Other minor variations are due to, e.g., unresolved seasonality patterns less amenable to our multiple least-square regression approach. 

\section{Discussion \& Conclusion}
	
The strong divergence between raw and deconvolved impact decay profiles shown in Fig. \ref{Fig2}-a highlights the subtle interplay between trades, 
their information content and price dynamics. Once past trades and predictable returns are removed, our results suggest that 
i) the price dynamics can be decomposed into a ``mechanical'' impact, identical for all trades (as expected on anonymous markets) and an information part; 
ii) the mechanical impact $I(\tau)$ of an isolated meta-order appears to be fully {\it transient} and decays all the way to zero  (Fig.\ref{Fig2}-a,b) \cite{Blanc}. 
We therefore believe that the previously reported asymptotic non-zero components are likely to
be caused by the impossibility to adequately subtract these components from the data -- echoing previous research \cite{Gomes2013}. 
Since our results on different time periods and markets are consistent (within error-bars) with a complete decay, the relevance of the simple fair-price argument proposed in \cite{Farmer:2011} is not clear to us (as also highlighted by
our toy-model above).


There are still a number of open questions that future research, both theoretical and empirical, should try to address. In particular, our approach explicitly assumes impact decay to be a quasi-linear time-invariant process (Eq. (\ref{MultiReg})), which is in fact Gatheral's model in discrete time \cite{Gatheral} (see also \cite{Bouchaud:2004}). This is a seemingly plausible assumption given the consistency of the results presented so far, but in contradiction with the square-root law itself, which suggests that impact is {\it not} additive. We have seen in our data set hints that our linear assumption is indeed not fully satisfactory. For
example, the impact decay $I(\tau)$ measured on the sub-sample of trades that correspond to the top-half participation rate over the last ten days is noticeably different from the one measured on the bottom half (the former decaying
faster than the latter). This of course should not be the case within the above quasi-linear description. Furthermore, concavity and time decay are not easy to disentangle. For example, if aggregated meta-orders were still in the concave square-root regime over five days, this would lead to an {\it apparent} decay of $\approx 60 \%$ of impact when fitted by a quasi-linear model, even if impact was fully permanent \cite{Benichou}. We have however checked that our results are essentially unchanged when regrouping two successive days in a new elementary time unit and redoing the above analysis, suggesting that time decay indeed dominates over persistent concavity for $\tau > 1$ day. Still,  
the correct way to marry the time dependency of impact with its intrinsically non-linear nature is not trivial and should be better understood in order to reach a final conclusion. From a theoretical point of view, 
investigating further latent liquidity models \cite{Toth:2011,Iacopo:2013} may allow one to go beyond the above quasi-linear framework.

\vskip 0.5cm

Acknowledgements: We thank R. B\'enichou, P. Blanc, J. De Lataillade, J. Donier, Z. Eisler, D. Farmer, J. Gatheral, S. Hardiman, E. Henry, L. Laloux, C.-A. Lehalle, F. Lillo, 
Y. Lemp\'eri\`ere, I. Mastromatteo, M. Potters, D. Taranto, B. T\'oth, 
H. Waelbroeck \& E. Zarinelli for many useful conversations on these issues.

\end{document}